\documentclass[english]{article}
\usepackage[T1]{fontenc}
\usepackage[latin9]{inputenc}
\usepackage{color}
\usepackage{textcomp}
\usepackage{amstext}
\usepackage{amssymb}
\usepackage{amsmath}
\usepackage{esint}

\RequirePackage{lineno}
\makeatletter

\newcommand{\lyxmathsym}[1]{\ifmmode\begingroup\def\b@ld{bold}
  \text{\ifx\math@version\b@ld\bfseries\fi#1}\endgroup\else#1\fi}

\makeatother

\usepackage{babel}
\begin{document}

\title{\textbf{Initiating the effective unification of black hole horizon area and entropy quantization with quasi-normal modes}}

\author{\textbf{$^{1*}$C. Corda, $^{2,3,**}$S. H. Hendi, $^{4,+}$R. Katebi, $^{5,++}$ N. O. Schmidt}}

\maketitle
\begin{center}
\textbf{$^{1}$}Istituto Universitario di Ricerca \textquotedbl{}Santa Rita\textquotedbl{},
Villa il Ventaglio (G.C.) via delle Forbici 24/26 - 50133 Firenze,
Italy; Institute for Theoretical Physics and Advanced Mathematics
Einstein-Galilei (IFM), Via Santa Gonda 14, 59100 Prato, Italy; International
Institute for Applicable Mathematics and Information Sciences (IIAMIS),
Hyderabad, India and Udine, Italy

\par\end{center}

\begin{center}
\textbf{$^{2}$}Physics Department and Biruni Observatory, College
of Sciences, Shiraz University, Shiraz 71454, Iran
\par\end{center}

\begin{center}
\textbf{$^{3}$}Research Institute for Astrophysics and Astronomy of Maragha (RIAAM)
P.O. Box 55134-441, Maragha, Iran
\par\end{center}

\begin{center}
\textbf{$^{4}$}Department of Physics, California State University Fullerton, 800 North State College Boulevard, Fullerton, CA 92831, USA
\par\end{center}

\begin{center}
\textbf{$^{5}$}Department of Mathematics, Boise State University,
1910 University Drive, Boise, ID, 83725, USA
\par\end{center}

\begin{center}
\textit{E-mail addresses:} \textcolor{blue}{$^{*}$cordac.galilei@gmail.com,
$^{**}$hendi@shirazu.ac.ir, $^{+}$rkatebi.gravity@gmail.com, $^{++}$nathanschmidt@u.boisestate.edu }
\par\end{center}

\begin{abstract}
Black hole (BH) area quantization may be the key to unlocking a unifying theory of quantum gravity (QG). Surmounting evidence in the field of BH research continues to support a horizon (surface) area with a discrete and uniformly spaced spectrum, but there is still no general agreement on the level spacing. In the specialized and important BH case study, our objective is to report and examine the pertinent groundbreaking work of the \emph{strictly thermal} and \emph{non-strictly thermal} spectrum level spacing of the BH horizon area quantization with included entropy calculations, which aims to tackle this gigantic problem. In particular, such work exemplifies a series of imperative corrections that eventually permits a BH's horizon area spectrum to be \emph{generalized} from strictly thermal to non-strictly thermal with entropy results, thereby capturing multiple preceding developments by launching an effective unification between them. Moreover, the results are significant because quasi-normal modes (QNM) and ``effective states'' characterize the transitions between the established levels of the non-strictly thermal spectrum.
\end{abstract}

\section{Introduction}
BHs are mighty creatures that generate chaos in space-time physics. In general, the laws of classical and modern physics break down when attempts are made to rigorously characterize the behavior of BHs and their effects. In order to advance science, fundamental problems such as the BH information paradox and event horizon firewalls \cite{Hawking2014, Firewall2014, JoshiNarayan2014, CordaNoInfoLoss} must be understood and nullified so the physical laws can be ``upgraded'' via the scientific method and tested in laboratory experiments \cite{infoparadox2013}.

There is a vast array of modern attacks that aim to conquer BHs by establishing a unified field theory with a new set of physical laws. Among these approaches, numerous mainstream unification candidates (and variations of them) exist, including, super-string theory \cite{stringtheorybook}, QG, loop quantum gravity (LQG) \cite{LQGquantization, LQGrelate, LQGrelate2, LQGrelate3, LQGquant2009}, Chern-Simons theory \cite{su2cht2010}, Yukawa $SO(10)$ theory \cite{yukawa2013}, E8 theory \cite{E8}, and others. Frequently, components and ideas from different theories are combined, adjusted, and ``hacked'' together (i.e. copy-and-paste methods) to establish new hybrid theoretical frameworks with customized capabilities, such as semi-classical physics, which intertwines aspects of quantum mechanics and classical mechanics. Currently, none of these candidates are accepted to be complete by mainstream science. For example, some frameworks like super-string theory \cite{stringtheorybook}, Yukawa $SO(10)$ theory \cite{yukawa2013}, and E8 theory \cite{E8} are incomplete because they require more spatial degrees of freedom to operate than 4D space-time can offer so they cannot be tested in the laboratory, while other theories are incomplete because they fail to fully describe paradoxical phenomena like BHs, which remain imposing, elusive, and continue to violate the modern laws of physics. Hence, the theories must be subjected to additional stringent scientific research, scrutiny,  debate, and experimentation so they can continue to evolve and achieve improved representational capabilities.

In this review paper, we focus on the surface area and entropy quantization of BH event horizons, where we identify and examine some key points, issues, and corrections in a chronological narrative of \textit{strictly thermal} and \textit{non-strictly thermal} results. Years ago, BH emissions and absorptions were only partially understood in terms of thermal energy and Hawking radiation \cite{Hawking1975}---and not all energy is heat energy---so BHs then were still relatively murky and restricted in this sense. However, in more recent years, the trailblazing work by the Parikh-Wilczek team \cite{Parikh-Wilczek2000} ignited a revolution in BH physics because they hacked the formula structure of the strictly thermal tunneling rate and exploited it to initiate a non-strictly thermal picture of BHs that encompasses \textit{all} energy, thereby circumventing numerous physical restrictions to further pave the way towards unification. So for this assignment, we discuss the finer points of this key generalization and its impact on BH area and entropy, where the pertinent, groundbreaking works of numerous, additional research teams are investigated. As mentioned above, there is a diverse landscape of candidate unification theories that may be applied to this particular BH aspect. Thus, from among the said candidates, we've selected a \textit{semi-classical} platform to launch a probe of BHs that exemplifies the underlying QG theory. For this work, we prefer this semi-classical, QG-based approach over existing unification candidates such as super-string theory \cite{stringtheorybook}, Yukawa $SO(10)$ theory \cite{yukawa2013}, and E8 theory \cite{E8} because their gravitational treatment adds too many spatial degrees of freedom. Moreover, a popular alternative to QG is LQG \cite{LQGquantization, LQGrelate, LQGrelate2, LQGrelate3, LQGquant2009}, which does have 4D space-time gravity built-in by default because it fundamentally operates on the principles of general relativity. Moreover, recent emerging LQG-based approaches do yield promising results for directly counting physical states with new quantization techniques and connections to semi-classical Bekenstein-Hawking entropy \cite{LQGquantization, LQGrelate, LQGrelate2, LQGrelate3, LQGquant2009}. For example we refer the reader to the holography \cite{holoSuss,holoEntropy}, the origin of thermodynamics \cite{originEntropy,originEntropy2,originEntropy3}, spin foams \cite{spinfoamEntropy, spinfoamEntropy2}, and the many open questions concerning the classical limit of generalizing static uncharged BHs to encompass charged, rotating cases. 

%
%

\section{Strictly thermal horizon area and entropy quantization} \label{sec:thermal}
First, we focus on Hawking temperature and review imperative results concerning the BH area and entropy quantization for a \textit{strictly thermal} spectrum and mass-energy level structure.

\subsection{Initial horizon area quantization boundaries} \label{sec:thermal:initiating}
In the early 1970s, Bekenstein \cite{Bekenstein1972, Bekenstein1973} observed that the (non-extremal) BH horizon area behaves as a classical adiabatic invariant and therefore conjectured that it should exemplify a \emph{discrete} eigenvalue spectrum with quantum transitions \cite{Hod1999,Hod2008}. To date, a major objective in BH physics research is to determine the unique spacing between the BH horizon area levels because surmounting scientific evidence seems to indicate that the BH horizon area spectrum is in fact quantized and uniformly distributed \cite{Hod1999, Hod2008}. Thus, our investigation launches from the \emph{particle} platform of wave-particle duality.

When a BH captures or releases a massive point particle, then the BH's mass unavoidably increases or decreases, respectively, which directly influences its horizon area \cite{Hod1999,Hod2008}. For the BH's \textit{uncharged} particle absorption process, it was ascertained \cite{Bekenstein1973} from Ehrenfest's theorem that the particle's center of mass must follow a classical trajectory and therefore it was demonstrated that the BH horizon area increase lower-bound is \cite{Hod1999,Hod2008}
\begin{equation}
 \label{eq:area-increase-lower-bound-classical}
 \Delta A = 8 \pi \mu b,
\end{equation}
where $\Delta A$ is the BH horizon area change, $\mu$ is the particle rest mass, and $b$ is the particle finite proper radius. In a quantum theory, Heisenberg's uncertainty principle applies to relativistic quantized particles \cite{Hod1999,Hod2008}---specifically, the radial position for the particle's center of mass is subject to an uncertainty of $b \ge \hbar/\mu$ because it cannot be localized with a degree of precision that supercedes its own Compton wavelength \cite{Hod1999,Hod2008}. Thus, for the uncharged particle absorption process, the uncertainty principle is the physical mechanism which defines the uncharged BH horizon area increase lower-bound as \cite{Hod1999,Hod2008}
\begin{equation}
 \label{eq:area-increase-lower-bound-uncharged-quantum}
\Delta A = 8 \pi l_p^2,
\end{equation}
where $l_p = \sqrt{\frac{G \hbar}{c^3}}$ is the Planck length in gravitational units $G = c = 1$. However, for the BH's \emph{charged} particle absorption process the ``uncertainty principle mechanism'' must be supplemented by a secondary physical mechanism---a Schwinger discharge for the BH vacuum polarization process \cite{Hod1999,Hod2008}. Hence, for the charged case, this ``vacuum polarization mechanism'' lets one bypass the reversible limit constraint and defines the charged BH horizon area increase lower-bound as \cite{Hod1999,Hod2008}
\begin{equation}
\label{eq:area-increase-lower-bound-charged-quantum}
 \Delta A = 4 \ln e l_p^2 = 4 l_p^2.
\end{equation}
Here, the lower-bounds of eqs. (\ref{eq:area-increase-lower-bound-uncharged-quantum}--\ref{eq:area-increase-lower-bound-charged-quantum}) are fully consistent with the analysis of \cite{vagenasEPL2010, vagenasPRL2010}.

Thus, as soon as one introduces quantum implications into the absorption process it becomes evident that eqs. (\ref{eq:area-increase-lower-bound-uncharged-quantum}--\ref{eq:area-increase-lower-bound-charged-quantum}) are in fact universal lower-bounds because they are independent of the BH parameters \cite{Hod1999,Hod2008}; this fundamental lower-bound's \emph{universality} strongly favors a uniformly-spaced quantum BH horizon area spectrum \cite{Hod1999,Hod2008}. Moreover, it is striking that although the results of eqs. (\ref{eq:area-increase-lower-bound-uncharged-quantum}--\ref{eq:area-increase-lower-bound-charged-quantum}) emerge from two distinct physical mechanisms, they are clearly of the same magnitude order \cite{Hod1999,Hod2008} and differ by a factor of $2\pi$ due to the existence of charge, which is further realized in \cite{vagenasEPL2010, vagenasPRL2010}. Hence, it was concluded that the BH horizon area quantization condition is of the form \cite{Hod1999,Hod2008}
\begin{equation}
\label{eq:horizon-area-quantization-spectrum}
 A_n = \gamma n l_p^2 \ \ ; \ \ n = 1,2,... \ \,
\end{equation}
where $\gamma$ is a dimensionless constant.

In \cite{Hod1999,Hod2008}, it was recognized that the exact values of eqs. (\ref{eq:area-increase-lower-bound-uncharged-quantum}--\ref{eq:area-increase-lower-bound-charged-quantum}) can be challenged because they operate on the assertion that the smallest possible particle radius is precisely equal to its Compton wavelength and because the particle size is inherently fuzzy. But it is clear that the $\gamma$ in both eqs. (\ref{eq:area-increase-lower-bound-uncharged-quantum}--\ref{eq:area-increase-lower-bound-charged-quantum}) cases must be of the magnitude order $\gamma \sim 4$ \cite{Hod1999,Hod2008}. Moreover, that ``the small uncertainty in the value of $\gamma$ is the price we must pay for not giving our problem a full quantum treatment'' \cite{Hod1999,Hod2008}. Therefore, the quantum analysis \cite{Hod1999,Hod2008} shifts from \emph{discrete particles} to \emph{continuous waves} due to the uncertainty of $\gamma$; this is allowed because of nature's wave-particle \emph{duality} of mass-energy---one must be able to infer the wave results from the particle results, and conversely. Consequently, the QNMs authorize one to explore BH perturbations from the perspective of such waves \cite{Hod1999, Hod2008, Maggiore2008, Corda-IJMPD-2012}. Specifically, QNMs enable one to characterize a BH's free oscillations, where the behavior of the radiated perturbations is reminiscent to the last pure dying tones of a ringing bell because the QNM frequencies are representative of the BH itself \cite{Hod1999,Hod2008, Maggiore2008, Corda-IJMPD-2012}. The perturbation field QNM states encode the scattering amplitude's pole singularities in the BH background \cite{Hod2008}. More specifically, the quantized states of the perturbation fields outside the BH are encoded with complex numbers for QNMs, where the BH perturbation fields transition between states in the ``BH perturbation field state space'' over ``state time''. The BH states of such complex-valued QNMs are equipped with the amplitude, real, and imaginary components. In BH physics and thermodynamics, it is imperative to be able to encode such QNM states and transitions for determining the asymptotic behavior of BH ringing frequencies---this is a challenging physical encoding problem that requires a proper, rigorous quantum treatment in order to further demystify and generalize the horizon area results of eqs. (\ref{eq:area-increase-lower-bound-classical}--\ref{eq:horizon-area-quantization-spectrum}).

\subsection{Perturbation field quasi-normal mode states} \label{sec:thermal:qnms}
To attack the massively complex encoding problem in Hawking's \textit{strictly thermal} radiation spectrum, Maggiore \cite{Maggiore2008} went on to demonstrate that the behavior of the BH perturbation field QNM states is identical to that of damped harmonic oscillators whose real frequencies are encoded as the 2D polar amplitude
\begin{equation}
\label{eq:amplitude-pythagorean}
 |\omega| = \sqrt{\omega_{\mathbb{R}}^2 + \omega_{\mathbb{I}}^2},
\end{equation}
rather than just $\omega_{\mathbb{R}}$, such that
\begin{equation}
\label{eq:pythagorean-components}
  \omega_{\mathbb{R}} = \sqrt{|\omega|^2 - \left(\frac{K}{2}\right)^2} \ \ \text{and} \ \ \omega_{\mathbb{I}} = \frac{K}{2}
\end{equation}
are the 2D Cartesian real and imaginary components, respectively, where $K$ is the damping coefficient. In eqs. (\ref{eq:amplitude-pythagorean}--\ref{eq:pythagorean-components}), the case $|\omega| = \omega_{\mathbb{R}}$ for $\omega_{\mathbb{I}} \ll |\omega|$ corresponds to lowly-excited, very long-lived perturbation states, whereas the ``opposite'' limit case $|\omega| = \omega_{\mathbb{I}}$ for $\omega_{\mathbb{R}} \ll \omega_{\mathbb{I}}$ corresponds to highly-excited, very short-lived perturbation states \cite{Maggiore2008}---so $|\omega| \simeq \omega_{\mathbb{I}}$ rather than $|\omega| \simeq \omega_{\mathbb{R}}$. The results of eqs. (\ref{eq:amplitude-pythagorean}--\ref{eq:pythagorean-components}) exemplify the three distinct QNM components---$|\omega|$, $\omega_{\mathbb{R}}$, and $\omega_{\mathbb{I}}$---that comply with Pythagorean's theorem of triangles for the precise determination of physical properties. 

In order to study the transition frequencies, one may use the Bohr's correspondence principle, which was published in 1923, to establish order in the chaos. In \cite{Hod1999,Hod2008}, it has been shown that transition frequencies at large quantum numbers should equal classical oscillation frequencies. Thus, the analysis \cite{Hod1999,Hod2008} focused on the ringing frequencies asymptotic behavior for the $n\rightarrow \infty$ limit, which are classified as highly-damped BH perturbation field QNM frequencies that operate under the assertion that such quantum transitions between states are \textit{instantaneous}. The transitions do not require time because it was established that $\omega = \omega_{\mathbb{R}} - i \omega_{\mathbb{I}}$ \cite{Hod1999,Hod2008}, such that $\tau \equiv \omega_{\mathbb{I}}^{-1}$ is the effective relaxation time which is required for the BH to return to a state of equilibrium, where $\tau$ is arbitrarily small as $n\rightarrow \infty$. On one hand, for each value of the angular momentum quantum number $l$, there exists an infinite number of QNMs for $n = 0,1,2,...$ with decreasing relaxation times (so the value of $\omega_{\mathbb{I}}$ increases) \cite{Hod1999,Hod2008}. On the other hand, $\omega_{\mathbb{R}}$ approaches a constant value as $n$ is increased \cite{Hod1999,Hod2008}. Hence, the amplitude of eq. (\ref{eq:amplitude-pythagorean}) is re-written for large $n$ as
\begin{equation}
\label{eq:amplitude-pythagorean-n}
 |\omega_n| = \sqrt{\omega_{n_{\mathbb{R}}}^2 + \omega_{n_{\mathbb{I}}}^2},
\end{equation}
which exhibits a BH energy level structure that is physically very reasonable, because both the amplitude component $|\omega_n|$ and the imaginary component $\omega_{n_{\mathbb{I}}}$ increase monotonically with the overtone number $n$ \cite{Maggiore2008}. Thus, the context of equivalent harmonic oscillators, $n=1$ is the least damped state for the lowest value of $|\omega|$, while $|\omega_n|$ is the larger state with a shorter lifetime \cite{Maggiore2008}. The asymptotic behavior of the highly-damped states is difficult to determine because of the effect of exponential divergence of the QNM eigenfunctions at the physical boundary of purely outgoing waves at the tortoise radial coordinate $r_{*}\rightarrow\infty$ \cite{Hod1999,Hod2008}. However, it is known for the simplest case of a Schwarzschild BH (SBH) with mass $M$ as \cite{Hod1999,Hod2008}
\begin{equation}
\label{eq:highly-damped-qnm-frequency}
M\omega_{n}=0.0437123-\frac{i}{4}(n+\frac{1}{2})+O[(n+1)^{-1/2}],
\end{equation}
a characteristic of the BH itself (in the $n \gg 1$ limit), which is only dependent upon $M$ and is \emph{independent} of $l$ and $\sigma$.

Moreover, it was shown in \cite{Hod1999,Hod2008} that the numerical limit $Re(M\omega_{n})\rightarrow0.0437123$ (as$\hspace{2mm} n \rightarrow\infty)$ agrees with the quantity $\ln3/(8\pi)$ and is thereby supported by thermodynamic and statistical physics. So when equipped with $\Delta A = 4\ln3 l_{p}^{2} $ from $A(M)=16\pi M^{2}$ and $dM=E=\hbar\omega$, one can identify
\begin{equation}
 \gamma_{Hod}(3)=4\ln3
\end{equation}
 for the quantum SBH horizon area spectrum of eq. (\ref{eq:horizon-area-quantization-spectrum}), which is upgraded to \cite{Hod1999,Hod2008}
\begin{equation}
\label{eq:horizon-area-quantization-spectrum2}
A_{n} = \gamma_{Hod}(3) l_{p}^{2} n.
\end{equation}
So the wave analysis is consistent with the particle analysis of the magnitude order $\gamma \sim 4$ \cite{Hod1999,Hod2008}; this result supports the wave-particle duality of mass-energy with an exactitude of mechanics, rather than statistics. From the statistical standpoint, eq. (\ref{eq:horizon-area-quantization-spectrum2}) is paramount because it complies with the semi-classical version of Christodoulou's reversible process, which is mechanistic in nature, and is independent of the thermodynamic relation between the BH horizon area $A_n$ and entropy $S_{BH}(n)$ \cite{Hod1999,Hod2008}. The accepted relation between $A_n$ and $S_{BH}(n)$ is pertinent if, for any $n$, the constraint
\begin{equation}
 \label{eq:spectrum-spacing}
\gamma_{Hod}(k)= 4 \ln k \hspace{2.5mm}; \hspace{2.5mm} k=2,3,...\ \ ,
\end{equation}
is satisfied, such that $g(n) \equiv e^{S_{BH}(n)}$ is the degeneracy of the $n$th area eigenvalue \cite{Hod1999,Hod2008}. Hence, the first independent derivation of $k$ was established \cite{Hod1999,Hod2008}, which still requires additional contemplation because there is still no general agreement on the spectrum level spacing. But eq. (\ref{eq:spectrum-spacing}) is still the only expression that is consistent with both the area-entropy thermodynamic relation, statistical physics, and Bohr's correspondence principle \cite{Hod1999,Hod2008}.

The lower-bound universality of eqs. (\ref{eq:area-increase-lower-bound-uncharged-quantum}--\ref{eq:area-increase-lower-bound-charged-quantum}) and the entropy universality suggest that the area spectrum of eq. (\ref{eq:horizon-area-quantization-spectrum2}) is valid not only for SBHs, but more sophisticated physical structures such as Kerr BHs (KBH) and Kerr-Newman BHs (KNBH) \cite{Hod1999,Hod2008}. Moreover, an assumption was proposed regarding the asymptotic behavior of highly-damped QNMs of generic KNBHs \cite{Hod1999,Hod2008}. Upon considering the first law of BH thermodynamics \cite{Hod1999,Hod2008}
\begin{equation}
dM=\Theta(M,a,Q) dA(M)+\Omega dJ
\end{equation}
for
\begin{equation}
 \Theta(M,a,Q) = \frac{r_{+}(M,a,Q)-r_{-}(M,a,Q)}{4 A(M)}
\end{equation}
and
\begin{equation}
 \Omega(M,a) = \frac{4\pi a}{A(M)},
\end{equation}
where the KNBH inner and outer horizons are
\begin{equation}
\begin{array}{rcl}
 r_{+}(M,a,Q) &=& M + \sqrt{M^{2}-a^{2}-Q^{2}}\\
 \\
 r_{-}(M,a,Q) &=& M - \sqrt{M^{2}-a^{2}-Q^{2}},
 \end{array}
\end{equation}
such that $a=J/M$ is the KNBH angular momentum per unit mass, one can find \cite{Hod1999,Hod2008}
\begin{equation}
\omega_{{n}_{\mathbb{R}}} \rightarrow \Theta(M,a,Q) \gamma_{Hod}(3) + \Omega(M,a) m,
\end{equation}
where $n\rightarrow\infty$, such that $m$ is the perturbation field's azimuthal eigenvalue that corresponds to its phase.

Along this approach, for large $n$, the strictly thermal asymptotic behavior \cite{Maggiore2008} was employed
\begin{equation}
\label{eq:SBH-QNM-asumptotic-behavior}
8\pi M\omega_{n}= \frac{\omega_{n}}{8\pi M}=\ln 3 + 2\pi i (n+\frac{1}{2})+O[(n+1)^{-1/2}],
\end{equation}
for the Hawking temperature
\begin{equation}\label{eq:SBH-Hawking-temperature}
 T_H = \frac{\hbar}{8\pi M}
\end{equation}
to re-write eq. (\ref{eq:amplitude-pythagorean-n}) as
\begin{equation}
\label{eq:amplitude-pythagorean2}
 \hbar|\omega_n| = \sqrt{m_0^2 + p_n^2},
\end{equation}
for the underlying QNM Pythagorean components
\begin{equation}
\label{eq:pythagorean-components2}
  m_0 = \omega_{n_{\mathbb{R}}} = T_H \ln 3 \ \ \text{and} \ \ p_n = \omega_{n_{\mathbb{I}}} = 2\pi T_H \left(n + \frac{1}{2} \right).
\end{equation}

In the very large $n$ approximation, the leading term in the imaginary part of the complex frequencies in eq. (\ref{eq:SBH-QNM-asumptotic-behavior}) becomes dominant and spin independent, while, strictly speaking, eq. (\ref{eq:SBH-QNM-asumptotic-behavior}) works only for scalar (spin 0) and gravitational (spin 2) perturbations, see \cite{Maggiore2008} for details. In the $p_n$ of eq. (\ref{eq:pythagorean-components2}), recall that the $2\pi$ mathematically relates a circular radius to a circular circumference and is the difference between the uncharged and charged area quantization lower bounds of eqs. (\ref{eq:area-increase-lower-bound-uncharged-quantum}--\ref{eq:area-increase-lower-bound-charged-quantum}) that complies with \cite{vagenasEPL2010, vagenasPRL2010} so one could hypothesize that this intriguing $2\pi$ critical value may suggest a fundamental relationship to a circularly-symmetric or spherically-symmetric physical topology. The formulation of $p_n$ \cite{Maggiore2008} is fascinating because it harmonizes a quantized particle with \emph{antiperiodic} boundary conditions on a circle of circumference length
\begin{equation} \label{eq:circle-length}
 L = \frac{\hbar}{T_H(M)} = 8\pi M.
\end{equation}
At this point, preparations were made to re-examine some aspects of quantum BH physics by assuming the relevant frequencies are $|\omega_n|$, rather than $\omega_{n_{\mathbb{R}}}$ \cite{Maggiore2008}.

Next, in \cite{Maggiore2008} some important quantized spacing results for the discrete BH area spectrum were recalled. First, the conjecture of \cite{Bekenstein1973} was noted \cite{Maggiore2008}, which proposed that the level spacing is in quantized units of $l_p^2$ and thereby resulted in the SBH area quantum $\Delta A = 8\pi l_p^2$ of eq. (\ref{eq:area-increase-lower-bound-uncharged-quantum}) so we label $\gamma_{Bek} = 8\pi$ as Bekenstein's dimensionless constant. Second, Maggiore \cite{Maggiore2008} recognized that the results of \cite{Hod1999,Hod2008} revealed a similar quantization, but utilized the SBH QNM properties to discover the different numerical coefficient, namely $\Delta A = \gamma_{Hod}(3) l_p^2$ of eq. (\ref{eq:horizon-area-quantization-spectrum2}).

Although the hypothesis \cite{Hod1999,Hod2008} is exciting (primarily due to some possible connections with LQG), it still exhibits some complications \cite{Maggiore2008}. Additional analysis on the term $\gamma_{Hod}(3)$ with its origin in $\omega_{n_{\mathbb{R}}}$ for eq. (\ref{eq:SBH-QNM-asumptotic-behavior}) is in fact \emph{not universal} because it does not comply with charged and/or rotating BHs \cite{Maggiore2008}. For example, in the case of a KBH or KNBH with $a = J/M$, one finds that the large $n$ limit and the limit $a \rightarrow 0$ do not commute because if one first considers $a \rightarrow 0$, then $\omega_{n_{\mathbb{R}}}$ does not reduce to $\ln 3 / (8\pi M)$ and instead vanishes as $a^{1/3}$, which means that the area quantum becomes arbitrarily small if one gives the BH an infinitesimal rotation \cite{Maggiore2008}. Similarly, in the case of a Reissner-Nordstr\"{o}m BHs (RNBH) or KNBH, one finds that $\omega_{n_{\mathbb{R}}}$ changes discontinuously if the limits $Q \rightarrow 0$ and $n \rightarrow \infty$ are interchanged \cite{Maggiore2008}. Thereafter, a couple of additional exploits were pointed out in Hod's conjecture \cite{Hod1999,Hod2008}, so it was initially concluded that it ``does not reflect any intrinsic property of the BH, and the would-be area quantum vanishes in various instances'' and that its ``area quantization holds only for a transition from (or to) a BH in its fundamental state, while transitions among excited levels do not obey it'' \cite{Maggiore2008}. But, after additional scrutiny and venture \cite{Maggiore2008}, it was determined that all of the above complications are deleted when, in the conjecture of \cite{Hod1999,Hod2008}, one replaces $\omega_{n_{\mathbb{R}}}$ with $|\omega_n|$! For large $n$ and the transition $n \rightarrow n-1$, eq. (\ref{eq:SBH-QNM-asumptotic-behavior}) and $|\omega_n| \simeq \omega_{n_{\mathbb{I}}}$ yield the absorbed energy $\Delta M = \hbar [|\omega_n| - |\omega_{n-1}|] = \hbar/(4M)$, such that \cite{Maggiore2008}
\begin{equation}\label{eq:maggiore-spacing}
 \Delta A = 32 \pi M \Delta M = 8 \pi l_p^2,
\end{equation}
which complies with the old results of \cite{Bekenstein1973} because $\gamma_{Bek} = 8 \pi$. Thus, given the equal spacing for $|\omega_n|$ at large $n$, all other transitions require a larger energy; i.e. $n \rightarrow n - 2$ consumes about twice the energy \cite{Maggiore2008}. Even if one dares to extrapolate at low $n$  (where semi-classical reasoning may be destroyed), the non-vanishing $\Delta A$ of eq. (\ref{eq:maggiore-spacing}) remains consistent on that magnitude order \cite{Maggiore2008}. Therefore, the final results of \cite{Maggiore2008} concluded that the spacing of eq. (\ref{eq:maggiore-spacing}) indicates a \textit{consistent} SBH horizon area quantization, which implies that $l_p$ is the minimum magnitude order length for the existential and generalized uncertainty principle.

Consequently, in terms of BH entropy and micro-states, the work of \cite{Maggiore2008} determined that for large $n$, the horizon area quantum is $\Delta A = \gamma_{Bek} l_p^2$, such that $\gamma_{Bek} = 8 \pi$ of \cite{Bekenstein1973} replaces $\gamma_{Hod}(3) = 4 \ln 3$ of \cite{Hod1999,Hod2008}. Thus, the total horizon area must be of the form \cite{Maggiore2008}
\begin{equation}\label{eq:bekenstein-bh-area}
 A = N \Delta A = N \gamma l_p^2,
\end{equation}
where the area quanta number $N = A / \Delta A$ is an integer but is \emph{not} the same as the integer $n$ (which is used to label the BH perturbation field QNM states). Hence, the BH entropy is defined as \cite{Bekenstein1973, Maggiore2008}
\begin{equation}\label{eq:bekenstein-bh-entropy}
 S_{BH} = \frac{A}{\delta},
\end{equation}
where
\begin{equation} \label{eq:bekenstein-bh-entropy-delta}
\delta = 4 l_p^2
\end{equation}
agrees with the approach of \cite{vagenasEPL2010, vagenasPRL2010} and additionally the LQG approaches of \cite{LQGrelate2, LQGrelate3, LQGquant2009} to the same order of magnitude. Therefore, at level $N(M)$, it was expected that the number of possible BH micro-states (or ``BH micro-state space cardinality'') is \cite{Maggiore2008}
\begin{equation}
 g(N) \propto e^{A/ \delta} = e^{N \Delta A / \delta} = e^{\gamma N/4}.
\end{equation}
Subsequently, upon fixing the constant for $N=1$ in eq. (\ref{eq:bekenstein-bh-entropy}), there is only one micro-state in the state space, namely $g(N) = 1$, which gives \cite{Maggiore2008}
\begin{equation}
 g(N) = e^{(\gamma /4)(N-1)}.
\end{equation}
This operates under the required \emph{assumption} that $g(N)$ is an integer, which restricts $\gamma$ to in the form $\gamma_{Hod}(k)$ of eq. (\ref{eq:spectrum-spacing}), such that $k$ is an integer \cite{Maggiore2008}; the value $\gamma_{Hod}(3)$ is in the form of $\gamma_{Hod}(k)$ but the value $\gamma_{Bek}$ is not---$\gamma_{Bek}$ is only in the form of $\gamma_{Hod}(k)$ if $k = e^{2\pi}$ because
\begin{equation}
 \gamma_{Bek} = 8\pi = 4(2\pi) = 4(\ln k)
\end{equation}
holds for the periodicity $\ln k = 2\pi$ but clearly violates the ``$k$ must be an integer'' or ``$k$-constraint'' assertion---we also note that $\gamma_{Hod}(e) = 4 \ln e$ takes a similar form to $\gamma_{Hod}(k)$ but also violates the $k$-constraint.

These attempts to restrict $\gamma$ raise a number of objections \cite{Maggiore2008}. First, even in the trusted semi-classical framework, $N$ is gigantic, therefore $g(N)$ is the exponential of a colossal number \cite{Maggiore2008}. Even if the number of micro-states must be an integer, there is no hope that a semi-classical (or even a classical and statistical) calculation can identify this quantity with a precision of order one, which is requisite to distinguishing between an integer and non-integer result \cite{Maggiore2008}. Moreover, the above $g(N)$ expression assumes that the horizon area quantum $\Delta A$ is legal from large $N$ down to $N = 1$, where this semi-classical approximation is unwarranted \cite{Maggiore2008}. So although we see that eqs. (\ref{eq:amplitude-pythagorean2}--\ref{eq:pythagorean-components2}) determine equally spaced levels in the limit of highly-excited states, the level spacing for lowly-excited states are not equally spaced \cite{Maggiore2008}.

Thus, when the value $\gamma_{Bek}$ \cite{Bekenstein1973} was employed in $S_{BH}(M) = \gamma_{Bek}N(M)/4$, the result \cite{Maggiore2008}
\begin{equation}
\label{eq:bh-entropy-last}
 S_{BH} = 2\pi N + O(1)
\end{equation}
was discovered, such that $g(N) \propto e^{2\pi N(M)}$, for the leading order in the large $N$ limit. Basically, eq. (\ref{eq:bh-entropy-last}) gives a discrete spectrum which indicates that the entropy is an adiabatic invariant in accordance to Bohr's correspondance principle \cite{Maggiore2008}. All of this replicates the BH behavior and perturbation field states in terms of highly-damped harmonic oscillators whose real frequencies are the amplitude-modulus $|\omega_n|$ (instead of $\omega_{n_\mathbb{R}}$) for the area quantization $\Delta A = \gamma_{Bek} l_p^2$ (instead of $\Delta A = \gamma_{Hod}(3) l_p^2$). At this point, we also note that $\Delta A = \gamma_{Bek} l_p^2$ was also obtained in the alternative approach of \cite{vagenasPRL2011} \textit{without} the use of QNMs---another remarkable result that supports this development.

\section{Non-strictly thermal horizon area and entropy quantization} \label{sec:thermalandmore}
Second, we focus on corrections to the Hawking temperature and review additional significant results regarding the BH area and entropy quantization for a \textit{non-strictly thermal} spectrum and mass-energy level structure.

\subsection{Corrections to the Hawking temperature and Bekenstein-Hawking area and entropy law} \label{sec:thermalandmore:corrections}
Parikh and Wilczek \cite{Parikh-Wilczek2000} launched some outstanding corrections to the Hawking temperature by reverse engineering the formula structure of the semi-classical tunneling rate and deploying it to spark a non-strictly thermal picture of BHs based on a dynamical geometry. More specifically, they demonstrated that Hawking's radiation spectrum cannot be strictly thermal \cite{Parikh-Wilczek2000}, where such a non-strictly thermal character indicates that the BH spectrum is also non-strictly continuous. By taking into account the conservation of energy with an exact calculation of the action for a spherically-symmetric tunneling particle, the Parikh-Wilczek team defined a SBH's emission probability as \cite{Parikh-Wilczek2000}
\begin{equation}
 \Gamma_{PW} \sim \exp \left[ - \frac{\omega}{T_H} \left(1 - \frac{\omega}{2M} \right) \right]
\end{equation}
(in $G = c = k_b = \hbar = \frac{1}{4\pi \epsilon_0} = 1$ Planck units), which includes the new term $\frac{\omega}{2M}$ for the thermal deviation correction \cite{Parikh-Wilczek2000}.

Thereafter, given the results of Ref. \cite{Parikh-Wilczek2000} and associated works (see \cite{rbanerjee} and the references therein), Banerjee and Majhi gave an additional tunneling probability correction by considering the back reaction effect of the BH space-time metric \cite{rbanerjee}. In particular, they demonstrated that a SBH's tunneling probability can be written as \cite{rbanerjee}
\begin{equation} \label{eq:banerjee-tunneling}
\begin{array}{rcl}
 \Gamma_{BM} &\sim& \exp{\left[ -8 \pi M \omega + 4 \pi \alpha \left( \frac{2 M \omega}{M^2 + \alpha} \right) \right]}\\
 &=& \exp{ \left[ -\left( \frac{8 \pi M^3}{M^2 + \alpha} \right) \omega \right]} = \exp{\left[- \frac{\omega}{T_h} \right]}
 \end{array}
\end{equation}
from eq. (33) in Ref. \cite{rbanerjee}, where $\alpha = \frac{1}{8 \pi}$ is a dimensionless parameter (corresponding to the prefactor $-\frac{1}{2}$ of the QG calculations) and their revised Hawking temperature is \cite{rbanerjee}
\begin{equation}
 T_h = T_H \left(1 + \frac{\alpha}{M^2} \right),
\end{equation}
such that the new $\left(1 + \frac{\alpha}{M^2} \right)$ term is the correction due to the (one loop) back reaction with self-gravitation \cite{rbanerjee}. Here, the Banerjee-Majhi team expressed the corrected Bekenstein-Hawking entropy as \cite{rbanerjee}
\begin{equation} \label{eq:banerjee-entropy}
 \begin{array}{rcl}
  S_{BM} &=& \frac{A}{4} - 8 \pi \alpha \ln M - 64 \pi^2 \alpha^2 \left[ \frac{1}{A} - \frac{16\pi\alpha}{2A^2} + \frac{(16\pi\alpha)^2}{3A^3}  - ...\right]\\
  && + \ \text{const.(independent of} \ M\text{)}\\
  &=& S_{BH} - 4\pi \alpha \ln S_{BH} - \frac{16\pi^2 \alpha^2}{S_{BH}} \left[1 - \frac{4\pi\alpha}{2 S_{BH}} + \frac{(4\pi\alpha)^2}{3S_{BH}^2} - ... \right]\\
  && + \ \text{const.(independent of} \ M\text{)}\\
 \end{array}
\end{equation}
from eq. (28) in Ref. \cite{rbanerjee}, where the original entropy is $S_{BH} = 4\pi M^2 = \frac{A}{4}$ and the horizon area is $A = 16\pi M^2$. In eq. (\ref{eq:banerjee-entropy}) the non-leading corrections are identified as a series of inverse powers of $A$ (or $S_{BH}$) \cite{rbanerjee}. Moreover, the presented results of eqs. (\ref{eq:banerjee-tunneling}--\ref{eq:banerjee-entropy}) apply to a SBH but are general enough to encompass other cases as well \cite{rbanerjee}. 

Henceforth, the authors of Ref. \cite{rbanerjee} go beyond the semi-classical SBH approximation via the Hamilton-Jacobi method and implement the single quantized particle action corrections for additional such BH cases in a sequel paper \cite{rbanerjee2}. More precisely, in order to adjust a BH's Hawking temperature and Bekenstein-Hawking area and area law, they demonstrate that the selection of a simple proportionality constant reproduces the one loop back reaction effect in space-time via conformal field theory methods \cite{rbanerjee2}. For example, the Banerjee-Majhi team \cite{rbanerjee2} engaged the Hamilton-Jacobi method to re-express a SBH's Bekenstein-Hawking entropy of eq. (\ref{eq:banerjee-entropy}) as
\begin{equation}\label{eq:banerjee-entropy2}
 S_{BM} = S_{BH} + 4\pi \beta_1 \ln S_{BH} + \frac{16\pi^2 \beta_2}{S_{BH}} + ...
\end{equation}
from eq. (71) in Ref. \cite{rbanerjee2} by including additional quantum corrections and eliminating $A$, such that $\beta_i = \alpha^i$ from eq. (31) in Ref. \cite{rbanerjee2} and the revised $\alpha$ from eq. (34) in Ref. \cite{rbanerjee2} are both dimensionless parameters. Similarly, the Bekenstein-Hawking entropy for an anti-de Sitter SBH was also given in eq. (75) of Ref. \cite{rbanerjee} as
\begin{equation}\label{eq:banerjee-entropy2-antidesitter}
 S_{BM} = S_{BH} + 4\pi \beta_1 \ln S_{BH} + ...,
\end{equation}
where $\beta_1 = -\frac{1}{4\pi}$ for the leading order logarithmic correction was obtained via a statistical method \cite{rbanerjee2}. Also, we note that the back reaction semi-classical QG results of Refs. \cite{rbanerjee, rbanerjee2} are fully compliant with the self-consistent, spatially-isotropic perturbation corrections in de Sitter space-time for the one loop vacuum polarization of the Bunch-Davies vacuum state given by P\'{e}rez-Nadal \cite{gperez}, where a spatially flat Robertson-Walker space-time is driven by a cosmological constant that is non-conformally coupled to a massless scalar field.

Furthermore, the said Hamilton-Jacobi incursion \cite{rbanerjee2} is exercised in more recent investigations \cite{Majhi,Majhi2}, where additional modifications to the Hawking temperature and Bekenstein-Hawking area and entropy law are achieved: in Ref. \cite{Majhi} the results of Ref. \cite{rbanerjee2} are further applied to a scalar particle to examine the fermion tunneling of a Dirac particle as it is blasted into a BH's event horizon, whereas in Ref. \cite{Majhi2} the approach \cite{rbanerjee2} is also implemented and analyzed for the boson (photon) tunneling across a BH's event horizon, such that the coefficient of the leading order correction of entropy is related to the trace anomaly \cite{rbanerjee3, Banerjee4}. In both works \cite{Majhi,Majhi2}, the newer outcomes are consistent with those of the original loop back reaction effect \cite{rbanerjee, rbanerjee2}. 

In an independent but related approach for obtaining the area and entropy corrections for BHs in Ho\u{r}ava-Lifshitz gravity, Majhi deployed a density matrix to compute the radiation spectrum for a perfect black body in the semi-classical limit \cite{Majhi3}. In this analysis, the reported temperature is proportional to the surface gravity of a BH in general relativity and the first law of BH thermodynamics is utilized to define the entropy as \cite{Majhi3}
\begin{equation}
 S_{Majhi} = \frac{1}{4} \left(A - \frac{a\Omega_a}{\Lambda} \ln \frac{A}{A_0} \right)
\end{equation}
from eq. (20) in Ref. \cite{Majhi3}, where in Einstein space $\Lambda$ is the cosmological constant, $2a$ is the constant scalar curvature, $A = \Omega_a r_h^2$ is the horizon area of the horizon radius $r_h = \sqrt{\frac{a}{3}}$, $A_0$ is the integration constant of the length square dimension, and the coordinate component $\Omega_a$ is obtained from the Ho\u{r}ava-Lifshitz line element of eq. (1) in Ref. \cite{Majhi3}. Ultimately, the level spacing of the area and entropy are achieved, which are characterized in terms of QNMs \cite{Majhi3}. So on one hand, the results indicate an equispaced entropy spectrum even though the spacing value is not the same \cite{Majhi3}, whereas on the contrary, the level spacing of the BH's area spectrum is not equidistant because the BH's entropy is disproportional to its horizon area \cite{Majhi3}---in either case, both outcomes comply with the Einstein-Gauss-Bonnet theory \cite{Majhi3}. Such insights revealed in the work of Ref. \cite{Majhi3} set the stage for the upcoming QNM aspects of the effective state discussion in the next section.

\subsection{Perturbation field quasi-normal mode effective states} \label{sec:thermalandmore:effectivestates}
The striking corrections constructed by the Parikh-Wilczek team \cite{Parikh-Wilczek2000} not only generalize Hawking's radiation to a non-strictly thermal, non-strictly continuous BH spectrum, but also generate a natural correspondence between Hawking radiation and the BH perturbation field QNM states; this supports the idea that BHs result in highly-excited states in an underlying unitary QG theory \cite{Corda-IJMPD-2012, Corda-AnnPhys-2013}. Moreover, the strictly thermal spectrum deviation results of Ref. \cite{infoparadox2013} strongly suggested that single particle quantum mechanical approaches may be essential for finding potential solutions to the BH information puzzle. Here, in relation to all of this, we discuss the new, developing notions of effective temperature and effective state \cite{Corda-IJMPD-2012, Corda-AnnPhys-2013, CordaHendiKatebiSchmidt-JHEP-2013, CordaAHEP2013} because they reveal an important semi-classical QG characterization of BH area and entropy quantization in terms of perturbation field QNM states and transitions.

Thus, after a careful and extensive examination of the non-strictly thermal, non-strictly continuous BH energy spectrum and the spherically-symmetric particle tunneling results \cite{Parikh-Wilczek2000} by Corda \cite{Corda-IJMPD-2012, Corda-AnnPhys-2013}, the conventional Hawking temperature $T_H(M)$ of eq. (\ref{eq:SBH-Hawking-temperature}) was replaced by defining the SBH's effective temperature of eq. (3) in Ref. \cite{Corda-IJMPD-2012} as
\begin{equation}\label{eq:SBH-effective-temperature}
\begin{array}{rclclcl}
 T_{E_{SBH}}(M, -\omega) &=& \frac{2M}{2M + (-\omega)} T_H &=& \frac{1}{4\pi(2M + (-\omega))} &&\\
 &=& \frac{1}{8\pi M_E(M, -\omega)} &=& \frac{1}{2\pi R_{E_{SBH}}(M, -\omega)} &=& \frac{1}{\beta_{E_{SBH}}(M, -\omega)}
\end{array}
\end{equation}
for the emission of an uncharged particle with energy-frequency $\omega$ so the SBH contracts, where $M$ is the SBH's initial mass before the emission, $M - \omega$ is the SBH's final mass after the emission, $M_E$ is the SBH's effective mass defined by eq. (5) in Ref. \cite{Corda-IJMPD-2012} as
\begin{equation} \label{eq:SBH-effective-mass}
 M_E(M, -\omega) = M + \frac{-\omega}{2} = M - \frac{\omega}{2},
\end{equation}
$R_{E_{SBH}}$ is the SBH's effective horizon defined by eq. (5) in Ref. \cite{Corda-IJMPD-2012} as
\begin{equation} \label{eq:SBH-effective-horizon}
 R_{E_{SBH}}(M, -\omega) = 2 M_E(M, -\omega),
\end{equation}
and $\beta_E$ is the SBH's effective Botzmann factor defined in eq. (12) of \cite{Corda-AnnPhys-2013}. The new effective quantities $T_{E_{SBH}}$, $M_E$, $R_{E_{SBH}}$, and $\beta_{E_{SBH}}$ are average quantities which characterize the effective state of a discrete process rather than a continuous process \cite{Corda-IJMPD-2012, Corda-AnnPhys-2013}. Thus, for example, eqs. (\ref{eq:SBH-effective-temperature}--\ref{eq:SBH-effective-horizon}) indicate that the circular antiperiodic boundary conditions of eq. (\ref{eq:circle-length}) can be replaced with with the effective horizon circumference
\begin{equation} \label{eq:effective-circle-length}
\begin{array}{rclclcl}
 L_{E_{SBH}}(M, -\omega) &=& \frac{1}{T_{E_{SBH}}(M, -\omega)} &=& 8 \pi M_E(M, -\omega)\\
 &=& \beta_{E_{SBH}}(M, -\omega) &=& 4 \pi R_{E_{SBH}}(M, -\omega)\\
  &=& \frac{2 \pi}{\kappa_{E_{SBH}}(M, -\omega)},
 \end{array}
\end{equation}
which is simply the geometric equivalence of Boltzmann's effective physical quantity $\beta_{E_{SBH}}$, such that the fundamentally related $\kappa_{E_{SBH}}$ is the SBH's effective surface gravity. Subsequently, the results of eqs. (\ref{eq:SBH-effective-temperature}--\ref{eq:SBH-effective-horizon}) were instrumental in the establishment of two additional effective quantities \cite{Corda-AnnPhys-2013}: the SBH's effective line element from eq. (14) in Ref. \cite{Corda-AnnPhys-2013}
\begin{equation}\label{eq:effective-sbh-line-element}
 d s_{E_{SBH}}^2 = -\left(1- \frac{R_{E_{SBH}}(M, -\omega)}{r}\right)dt^2 + \frac{dr^2}{1 - \frac{R_{E_{SBH}}(M, -\omega)}{r}} + r^2(\sin^2 \theta d \phi^2 + d \theta^2),
\end{equation}
which encompasses the dynamical geometry of the SBH during the emission or absorption of the particle. Through a rigorous examination of Hawking's arguments \cite{Banerjee-MajhiPLB, Hawking1979}, the Euclidean form of eq. (18) in Ref. \cite{Corda-AnnPhys-2013} was successfully presented as
\begin{equation}\label{eq:effective-hawking-euclidean-form}
 d s_{E_{SBH}}^2 = x^2 \left[ \frac{d \tau}{4M\left(1 - \frac{\omega}{2M} \right)} \right]^2 + \left( \frac{r}{R_E(M, -\omega)} \right)^2 dx^2 + r^2(\sin^2 \theta d \phi^2 + d \theta^2),
\end{equation}
which is regular at $x = 0$ and $r = R_E(M, -\omega)$ and permits one to rigorously obtain eq. (\ref{eq:effective-sbh-line-element}). In Refs. \cite{Banerjee-MajhiPLB, Hawking1979} it was shown that $\tau$ serves as an angular variable with the periodicity of $\beta_{E_{SBH}} = L_{E_{SBH}}$ in eq. (\ref{eq:effective-circle-length}) with the underlying antiperiodic boundary conditions.

Henceforth, the procedure of Ref. \cite{Banerjee-Majhi2009} led to the corrected physical states for bosons and fermions from eq. (15) in Ref. \cite{Corda-AnnPhys-2013} as
\begin{equation}\label{eq:effective-fermion-and-boson-states}
\begin{array}{rclcl}
 | \Psi \rangle_{boson} &=& (1 - \exp(\frac{-\omega}{T_E(M, -\omega)})^{\frac{1}{2}} \Sigma_n \exp(-\omega 4 \pi n M_E(M, -\omega) | n_{out}^{Left} \rangle \\
 && \otimes \ | n_{out}^{Right} \rangle\\
\\
| \Psi \rangle_{fermion} &=& (1 + \exp(\frac{-\omega}{T_E(M, -\omega)})^{-\frac{1}{2}} \Sigma_n \exp(-\omega 4 \pi n M_E(M, -\omega) | n_{out}^{Left} \rangle \\
&& \otimes \ | n_{out}^{Right} \rangle,
\end{array}
\end{equation}
which respectively correspond to the emission probability distributions
\begin{equation}\label{eq:effective-fermion-and-boson-probability-distributions}
\begin{array}{rcl}
 \langle n \rangle_{boson} &=& \frac{1}{\exp(\frac{-\omega}{T_E(M, -\omega)}) -1}\\
 \\
\langle n \rangle_{fermion} &=& \frac{1}{\exp(\frac{-\omega}{T_E(M, -\omega)}) +1}.
\end{array}
\end{equation}
from eq. (16) in Ref. \cite{Corda-AnnPhys-2013}. At this point, we note that in order to compute the SBH effective parameters for the absorption of an uncharged particle with energy-frequency $\omega$, the $-\omega$ argument of eqs. (\ref{eq:SBH-effective-temperature}--\ref{eq:effective-hawking-euclidean-form}) may be quickly replaced with $+\omega$; if we wish to reference both emission and absorption simultaneously in such formulas, it is straightforward to specify $\pm \omega$.

Next, Corda's attack \cite{Corda-IJMPD-2012} deployed eq. (\ref{eq:SBH-effective-temperature}) to re-write eq. (\ref{eq:pythagorean-components2}) in the corrected form
\begin{equation} \label{eq:pythagorean-components2-emission2}
  m_n = T_E(M, -|\omega_n|) \ln 3 \ \ \text{and} \ \ p_n = T_E(M, -|\omega_n|) 2\pi i \left(n + \frac{1}{2} \right),
\end{equation}
which takes into account the non-strictly thermal behavior of the SBH, where
\begin{equation}\label{eq:SBH-QNM}
\omega_{n}=m_n + p_n + \mathcal{O}(n^{-\frac{1}{2}}).
\end{equation}
We stress that, although eqs. (\ref{eq:pythagorean-components2-emission2}) and (\ref{eq:SBH-QNM}) have only been intuitively derived \cite{Corda-IJMPD-2012}, they have been rigorously derived in the appendix of Ref. \cite{CordaEPJC}. In Ref. \cite{CordaEPJC} it has been also shown that in the very large $n$ approximation, the leading term in the imaginary part of the complex frequencies in eq. (\ref{eq:SBH-QNM}) becomes dominant and spin independent, while, strictly speaking, eq. (\ref{eq:SBH-QNM}) works only for scalar  and gravitational perturbations---see Ref. \cite{CordaEPJC} for details. Then, considering the leading term in the imaginary part of the complex frequencies, eq. (24) in Ref. \cite{Corda-IJMPD-2012} gives
\begin{equation} \label{eq:emission-qnm}
|\omega_{n}|=M-\sqrt{M^{2}-\frac{1}{2}(n+\frac{1}{2})}
\end{equation}
for emission. In eq. (\ref{eq:emission-qnm}) it was observed that the emission $n \rightarrow n -1$ gives the energy variation of eq. (29) in Ref. \cite{Corda-IJMPD-2012} as
\begin{equation}\label{eq:Corda-mass-change}
 \Delta M_n = |\omega_{n-1}| - |\omega_n| = -f(M,n)
\end{equation}
for the spacing of eq. (\ref{eq:maggiore-spacing}) as
\begin{equation}\label{eq:Corda-area-change}
 \Delta A_{SBH}(M, \Delta M_n) = 32 \pi M \Delta M_n = -32\pi M \times f(M,n) \approx -\gamma_{Bek}
\end{equation}
in the very large $n$ limit, which is the same order of magnitude as the original area quantization result \cite{Bekenstein1973}---the $f(M,n)$ of eqs. (\ref{eq:Corda-mass-change}--\ref{eq:Corda-area-change}) was constructed in eq. (30) of Ref. \cite{Corda-IJMPD-2012}. We recall that the SBH's horizon area $A_{SBH}$ is related to its mass $M$ via the relation $A_{SBH} = 16\pi M^2$ \cite{Bekenstein1973}. From this, one observes that if $A_{SBH}$ is quantized as $|\Delta A| = \gamma_{Bek}$ \cite{Bekenstein1973, Maggiore2008} and $|\Delta A| = \gamma_{Hod}(3)$ \cite{Hod1999, Hod2008}, then the SBH's total horizon area must be \cite{Corda-IJMPD-2012}
\begin{equation}\label{eq:corda-sbh-horizon-area}
 A_{SBH}(M,n) = N_{SBH}(M,n) |\Delta A_{SBH}(M,n)| = 16\pi M^2 = 4 \pi R_{SBH}^2,
\end{equation}
for the SBH's event horizon $R_{SBH} = 2M$, such that eq. (33) in Ref. \cite{Corda-IJMPD-2012} is
\begin{equation}\label{eq:corda-sbh-N}
 N_{SBH}(M,n) = \frac{A_{SBH}(M,n)}{|\Delta A_{SBH}(M,n)|} = \frac{16 \pi M^2}{32 \pi M \Delta M_n} = \frac{M}{2 f(M,n)},
\end{equation}\label{eq:Corda-sbh-N}
where the well-known SBH's Bekenstein-Hawking entropy \cite{Bekenstein1972, Bekenstein1973, Hawking1975} was re-written as \cite{Corda-IJMPD-2012}
\begin{equation}\label{eq:Corda-SBH-entropy}
\begin{array}{rcl}
 S_{SBH}(M,n) &=& \frac{A_{SBH}(M,n)}{4}\\
&=& 8 \pi N_{SBH}(M,n) M | \Delta M_n |\\
&=& 8 \pi N_{SBH}(M,n) M \times f(M,n),
\end{array}
\end{equation}
which indicates the crucial result that $S_{SBH}$ is a function of the quantum overtone number $n$ \cite{Corda-IJMPD-2012}.

On the other hand, it is a common and general belief that there is no reason to expect that the Bekenstein-Hawking entropy will be the whole answer for a correct unitary theory of QG \cite{Shankaranarayanan}. For a better understanding of a BH's entropy one needs to go beyond Bekenstein-Hawking entropy and identify the sub-leading corrections \cite{Shankaranarayanan}. Hence, the quantum tunneling approach can be used to obtain the sub-leading corrections to the second order approximation \cite{jzhang}, where one observes that the BH's entropy
\begin{equation} \label{eq: entropia totale}
S_{total}=S_{BH}-\ln S_{BH}+\frac{3}{2A} 
\end{equation}
contains three distinct parts: the usual Bekenstein-Hawking entropy, the logarithmic term and the inverse area term \cite{jzhang}. In fact, if one wants to satisfy the unitary QG theory, then the logarithmic and inverse area terms must be requested \cite{jzhang}. Note: the coefficient of the leading order correction depends on the nature of the theory. Apart from a coefficient, this correction to the BH's entropy is consistent with the one of LQG \cite{jzhang}, where the coefficient of the logarithmic term has been rigorously fixed at $\frac{1}{2}$ \cite{jzhang,gosh}. Therefore, the expression of eq. (\ref{eq:Corda-SBH-entropy}) for Bekenstein-Hawking entropy permits us to re-write eq. (\ref{eq: entropia totale})
as \cite{Corda-IJMPD-2012}
\begin{equation} \label{eq: entropia totale 2}
S_{{total}_{SBH}}=8\pi NM\cdot f(M,n)-\ln\left[8\pi NM\cdot f(M,n)\right]+\frac{3}{64\pi NM\cdot f(M,n)}.
\end{equation}

In the top line of eq. (\ref{eq:Corda-SBH-entropy}), observe that denominator $4$, which divides the numerator $A_{SBH}$ to compute the resulting $S_{SBH}$, is reminiscent of the  $\delta$ from Refs. \cite{Bekenstein1973, vagenasEPL2010} in eqs. (\ref{eq:bekenstein-bh-entropy}--\ref{eq:bekenstein-bh-entropy-delta}). Additionally, note that the results of eqs. (\ref{eq:Corda-area-change}) and (\ref{eq:Corda-SBH-entropy}) indicate that the SBH's Bekenstein-Hawking entropy change is
\begin{equation}
 \Delta S_{SBH}(M,n) = \frac{\Delta A_{SBH}(M,n)}{4},
\end{equation}
where clearly a change of \textit{negative entropy} ($\Delta S_{SBH} < 0$) recurs for absorption transitions because energy is conserved in 4D space-time.

Therefore, in order to incorporate the emerging SBH effective state framework, eqs. (\ref{eq:corda-sbh-horizon-area}--\ref{eq:Corda-SBH-entropy}) become
\begin{equation}\label{eq:corda-sbh-horizon-area-effective}
\begin{array}{rcl}
 A_{E_{SBH}}(M, \Delta M_n) &=& N_{E_{SBH}}(M, \Delta M_n) |\Delta A_{E_{SBH}}(M, \Delta M_n)|\\
 \\
 &=& 16\pi M_E^2(M, \Delta M_n) = 4 \pi R_{E_{SBH}}^2(M, \Delta M_n),
 \end{array}
\end{equation}
\begin{equation}\label{eq:corda-sbh-N-effective}
\begin{array}{rcl}
 N_{E_{SBH}}(M, \Delta M_n) &=& \frac{A_{E_{SBH}}(M, \Delta M_n)}{|\Delta A_{E_{SBH}}(M, \Delta M_n)|}\\
\\
 &=& \frac{16\pi M_E^2(M, \Delta M_n)}{32 \pi M_E(M, \Delta M_n) n \times f(M,n)}\\
\\
&=& \frac{M_E(M, \Delta M_n)}{2 f(M,n)},
\end{array}
\end{equation}
and
\begin{equation}\label{eq:Corda-SBH-entropy-effective}
\begin{array}{rcl}
 S_{E_{SBH}}(M, \Delta M_n) &=&  \frac{A_{E_{SBH}}(M, \Delta M_n)}{4} = \pi R_{E_{SBH}}^2(M, \Delta M_n) \\
 \\
 &=& 8 \pi N_{E_{SBH}}(M, \Delta M_n) M_E(M, \Delta M_n) | \Delta M | \\
\\
&=& 8 \pi N_{E_{SBH}}(M, \Delta M_n) M_E(M, \Delta M_n) \times f(M,n)\\
\\
&=& \frac{f(M,n)}{T_E(M, \Delta M_n)}.
\end{array}
\end{equation}
One also obtains the total effective entropy as
\begin{equation}
S_{total_{E_{SBH}}}(f(M,\Delta M_{n}))=\frac{f(M,n)}{T_{E}(M,\Delta M_{n})}-\ln\left[\frac{f(M,n)}{T_{E}(M,\Delta M_{n})})\right]+\frac{3T_{E}(M,\Delta M_{n})}{8f(M,n)}.\label{eq: entropia totale 3}
\end{equation}
Hence, the effective state quantities of eqs. (\ref{eq:corda-sbh-horizon-area-effective}--\ref{eq: entropia totale 3}) recognize the seemingly pertinent, disjoint aspects of the candidate horizon area theories of Bekenstein \cite{Bekenstein1972, Bekenstein1973}, Hod \cite{Hod1999, Hod2008}, and Maggiore \cite{Maggiore2008} by replacing Hawking's strictly thermal $T_H$ \cite{Hawking1975, Hawking1979} with the non-strictly thermal $T_E$ \cite{Corda-IJMPD-2012} to establish a preliminary generalization and unification.

Thereafter, subsequent work initiated an effective state framework generalization from SBHs \cite{Corda-IJMPD-2012} to KBHs \cite{CordaHendiKatebiSchmidt-JHEP-2013}, which was largely inspired by the discoveries of Refs. \cite{Medvel-Vagenas2005, Arzano-Medved-Vagenas2005, Vagenas2008, Medved2008}. It is known that the quantifiable difference between a SBH and a KBH is the angular momentum components \cite{CordaHendiKatebiSchmidt-JHEP-2013}. Hence, for this the the KBH's effective angular momentum as $J_E(M, \Delta M_n) = M_E(M, \Delta M_n) \alpha_E(M, \Delta M_n)$ \cite{CordaHendiKatebiSchmidt-JHEP-2013}, where the KBH's effective specific angular momentum from eq. (3.13) in Ref. \cite{CordaHendiKatebiSchmidt-JHEP-2013} is expressed as
\begin{equation} \label{eq:alpha-effective-angular-momentum}
 \alpha_E(M, \Delta M_n) = \frac{J_E(M, \Delta M_n)}{M_E(M, \Delta M_n)}
\end{equation}
for the additional KBH's effective angular momentum components
\begin{equation} \label{eq:delta-effective-angular-momentum}
  \Delta_E(M, \Delta M_n) = r^2 - 2 M_E(M, \Delta M_n) r + \alpha_E^2(M, \Delta M_n)
\end{equation}
and
\begin{equation} \label{eq:sigma-effective-angular-momentum}
  \Sigma_E(M, \Delta M_n) = r^2 + \alpha_E^2(M, \Delta M_n) \cos^2 \theta
\end{equation}
from eqs. (3.14--3.15) in Ref. \cite{CordaHendiKatebiSchmidt-JHEP-2013} that authorized the identification of the KBH's effective outer and inner horizons
\begin{equation} \label{eq:KBH-outer-and-inner-horizons-effective}
\begin{array}{rcl}
R_{{+E}_{KBH}}(M, \Delta M_n) &=& M_{E}(M, \Delta M_n)+\sqrt{M_{E}^{2}(M, \Delta M_n)-\alpha_{E}^{2}(M, \Delta M_n)}\\
\\
R_{{-E}_{KBH}}M, \Delta M_n) &=& M_{E}(M, \Delta M_n)-\sqrt{M_{E}^{2}(M, \Delta M_n)-\alpha_{E}^{2}(M, \Delta M_n)},
\end{array}
\end{equation}
and the corresponding KBH's effective line element
\begin{equation} \label{eq:KBH-line-element-effective}
\begin{array}{rcl}
ds_{E_{KBH}}^{2} &=&-\left(1-\frac{2M_{E}(M, \Delta M_n)r}{\Sigma_{E}(M, \Delta M_n)}\right)dt^{2}-\frac{4M_{E}(M, \Delta M_n)\alpha_{E}(M, \Delta M_n)r\sin^{2}\theta}{\Sigma_{E}(M, \Delta M_n)}dtd\phi\\
\\
&&+\frac{\Sigma_{E}(M, \Delta M_n)}{\triangle_{E}(M, \Delta M_n)}dr^{2} +\Sigma_{E}(M, \Delta M_n)d\theta^{2}\\
\\
&&+\left(r^{2}+\alpha_{E}^{2}(M, \Delta M_n) +2M_{E}(M, \Delta M_n)\alpha_{E}^{2}(M, \Delta M_n)r\sin^{2}\theta\right)\\ 
\\
&&\sin^{2}\theta d\phi^{2},
\end{array}
\end{equation}
respectively, which takes into due account the KBH's dynamical geometry as it emits or absorbs particles \cite{CordaHendiKatebiSchmidt-JHEP-2013}. From there, eqs. (\ref{eq:alpha-effective-angular-momentum}--\ref{eq:KBH-outer-and-inner-horizons-effective}) permitted the definition of the KBH's effective (outer) horizon area of eq. (3.19) in Ref. \cite{CordaHendiKatebiSchmidt-JHEP-2013} as
\begin{equation}\label{eq:KBH-effective-horizon-area}
 \begin{array}{rcl}
A_{{+E}_{KBH}}(M, \Delta M_n) = 4\pi\left(R_{{+E}_{KBH}}^{2}(M, \Delta M_n)+\alpha_{E}^{2}(M, \Delta M_n)\right)\\
\\
\ \ \ \ =8\pi \left( M_{E}^{2}(M, \Delta M_n) + \sqrt{M_{E}^{4}(M, \Delta M_n)-J_{E}^{2}(M, \Delta M_n)} \right),
\end{array}
\end{equation}
the KBH's effective temperature of eq. (3.20) in Ref. \cite{CordaHendiKatebiSchmidt-JHEP-2013} as
\begin{equation}\label{eq:KBH-effective-temperature}
 \begin{array}{l}
T_{{+E}_{KBH}}(M, \Delta M_n)=\frac{R_{{+E}_{KBH}}(M, \Delta M_n)-R_{{-E}_{KBH}}(M, \Delta M_n)}{A_{+E}(M, \Delta M_n)}\\
\\
\ \ \ \ =\frac{\sqrt{M_{E}^{4}(M, \Delta M_n)-J_{E}^{2}(M, \Delta M_n)}}{4\pi M_{E}(M, \Delta M_n) \left(M_{E}^{2}(M, \Delta M_n)+\sqrt{M_{E}^{4}(M, \Delta M_n)-J_{E}^{2}(M, \Delta M_n)}\right)}
\end{array}
\end{equation}
and the KBH's effective area quanta of eq. (3.22) in Ref. \cite{CordaHendiKatebiSchmidt-JHEP-2013} as
\begin{equation}\label{eq:KBH-effective-area-quanta}
 \Delta A_{{+E}_{KBH}}(M, \Delta M_n) = 16 \pi M_E(M, \Delta M_n) \left[1 + \left(1 - \frac{J_E^2(M, \Delta M_n)}{M_E^4(M, \Delta M_n)} \right)^{-\frac{1}{2}} \right] \Delta M_n
\end{equation}
for the KBH's effective area quanta number of eq. (3.23) in Ref. \cite{CordaHendiKatebiSchmidt-JHEP-2013} as
\begin{equation}\label{eq:KBH-effective-area-quanta-number}
N_{{+E}_{KBH}}(M, \Delta M_n) = \frac{A_{{+E}_{KBH}}(M, \Delta M_n)}{|\Delta A_{{+E}_{KBH}}(M, \Delta M_n)|} = \frac{M_E(M, \Delta M_n)}{2f(M,n)},
\end{equation}
which enabled the KBH's effective Bekenstein-Hawking entropy of eq. (3.24) in Ref. \cite{CordaHendiKatebiSchmidt-JHEP-2013} to be identified as
\begin{equation}\label{eq:KBH-effective-Bek-Hawk-entropy}
\begin{array}{rcl}
 S_{{+E}_{KBH}} (M, \Delta M_n) &=& \frac{A_{{+E}_{KBH}}(M, \Delta M_n)}{4} \\
\\
&=& 8 \pi N_{{+E}_{KBH}}(M, \Delta M_n) M_E(M, \Delta M_n) \times f(M,n).
\end{array}
\end{equation}
Thus, for $J_E \ll M_E^2$, it was confirmed in Ref. \cite{Corda-IJMPD-2012} that eqs. (\ref{eq:KBH-effective-horizon-area}--\ref{eq:KBH-effective-Bek-Hawk-entropy}) reduce to the SBH case of eqs. (\ref{eq:corda-sbh-horizon-area}--\ref{eq:Corda-SBH-entropy-effective}).

Consequently, following the QNM KBH effective state framework \cite{CordaHendiKatebiSchmidt-JHEP-2013}, the constructions were generalized to a \textit{non-extremal} RNBH version \cite{CordaAHEP2013}. For this implementation, a new definition of $\Delta M_n$ for RNBH QNMs was formulated to construct the new RNBH effective quantities \cite{CordaAHEP2013}. Starting from eq. (40) in Ref. \cite{CordaAHEP2013} the RNBH's effective charge was defined for small $Q$ as
\begin{equation}\label{eq:RNBH-effective-charge}
 Q_E(Q, q) = \frac{Q + (Q \pm q)}{2},
\end{equation}
where $Q$ is the RNBH's initial charge before the transition and $Q \pm q$ is the RNBH's final charge after the transition. The BH's $M_E$ of eq. (\ref{eq:SBH-effective-mass}) and the RNBH's $Q_E$ of eq. (\ref{eq:RNBH-effective-charge}) can be used to identify the RNBH's effective line element as
\begin{equation}
\begin{array}{rcl}
ds_{E_{RNBH}}^{2} &=& \left(1 - \frac{2 M_E(M, \Delta M)}{r} + \frac{Q_E^2(Q, q)}{r^2} \right)dt^2 - \frac{d r^2}{1 - \frac{2 M_E(M, \Delta M)}{r} + \frac{Q_E^2(Q, q)}{r^2}}  \\
\\
&&- r^2 d \theta^2 - r^2 \sin^2 \theta d \phi^2.
\end{array}
\end{equation}
Next, for a quantum transition between the levels $n$ and $n-1$, the RNBH QNM definition of $\Delta M_n$ in Ref. \cite{CordaAHEP2013} and the $Q_E$ of eq. (\ref{eq:RNBH-effective-charge}) were deployed to define the RNBH's effective outer and inner horizons from eq. (60) in Ref. \cite{CordaAHEP2013} as
\begin{equation}
\label{eq:RNBH-effective-outer-horizon-QNM}
\begin{array}{rcl}
 R_{{+E}_{RNBH}}(M, \Delta M_n, Q, q) &=& M_E(M, \Delta M_n) + \sqrt{M_E^2(M, \Delta M_n) - Q_E^2(Q, q)}\\
 \\
 R_{{-E}_{RNBH}}(M, \Delta M_n, Q, q) &=& M_E(M, \Delta M_n) - \sqrt{M_E^2(M, \Delta M_n) - Q_E^2(Q, q)},
 \end{array}
\end{equation}
the RNBH's effective (outer) horizon area as \cite{CordaAHEP2013}
\begin{equation} \label{eq:RNBH-effective-outer-horizon-area-QNM}
\begin{array}{l}
A_{{+E}_{RNBH}}(M,\Delta M_n,Q,q) = 4 \pi R_{{+E}_{RNBH}}^2(M, \Delta M_n, Q, q) \\
\\
\ \ \ \ \ \ \ \ = 4 \pi \left(M_E(M, \Delta M_n) + \sqrt{M_E^2(M, \Delta M_n) - Q_E^2(Q, q)}\right)^2,
\end{array}
\end{equation}
the RNBH's effective horizon area change as \cite{CordaAHEP2013}
\begin{equation}
 \label{eq:RNBH-effective-horizon-area-change-QNM}
 \Delta A_{{+E}_{RNBH}}(M,\Delta M_n,Q,q) = \frac{2 \Delta M_n q + \pi Q^3}{(M^2 - Q^2)^{3/2}},
\end{equation}
the RNBH's effective Bekenstein-Hawking entropy as \cite{CordaAHEP2013}
\begin{equation} \label{eq:RNBH-effective-outer-entropy-QNM}
 S_{{+E}_{RNBH}}(M, \Delta M_n, Q, q) = \frac{A_{{+E}_{RNBH}}(M,\Delta M_n,Q,q)}{4},
\end{equation}
the RNBH's effective Bekenstein-Hawking entropy change as \cite{CordaAHEP2013}
\begin{equation}
  \label{eq:RNBH-effective-entropy-change-QNM}
 \Delta S_{{+E}_{RNBH}}(M,\Delta M_n,Q,q) = \frac{\Delta A_{{+E}_{RNBH}}(M,\Delta M_n,Q,q)}{4},
\end{equation}
and the RNBH's effective quantum area number as \cite{CordaAHEP2013}
\begin{equation}
 \label{eq:RNBH-effective-outer-area-quanta-number-QNM}
 N_{{+E}_{RNBH}}(M,\Delta M_n,Q,q) = \frac{A_{{+E}_{RNBH}}(M,\Delta M_n,Q,q)}{|\Delta A_{{+E}_{RNBH}}(M,\Delta M_n,Q,q)|}.
\end{equation}
Thus, for $Q_E^2 \ll M_E^2$, it was confirmed in Ref. \cite{CordaAHEP2013} that eqs. (\ref{eq:RNBH-effective-outer-horizon-area-QNM}), (\ref{eq:RNBH-effective-outer-entropy-QNM}), and (\ref{eq:RNBH-effective-outer-area-quanta-number-QNM}) reduce to the corresponding effective quantities of the SBH case for eqs. (\ref{eq:corda-sbh-horizon-area-effective}--\ref{eq:Corda-SBH-entropy-effective}).

\section{Conclusion} \label{sec:conclusion}
In this review paper, we reported and examined the pertinent groundbreaking work of the strictly thermal and non-strictly thermal spectrum level spacing of the BH horizon area and entropy quantization from a semi-classical QG approach. For this, we chronologically reviewed a series of imperative corrections that eventually permits the Hawking radiation and the Bekenstein-Hawking horizon area and entropy spectrum to be generalized from strictly thermal to non-strictly thermal with QNMs and effective states \cite{Corda-IJMPD-2012, Corda-AnnPhys-2013, CordaHendiKatebiSchmidt-JHEP-2013, CordaAHEP2013}, which are significant because they further exemplify the underlying QG theory. In general, all of the works presented in this review are important to physics because the characteristic physical laws of BHs must be understood in order to resolve, for example, the puzzles imposed by the BH information paradox and firewalls \cite{Hawking2014, Firewall2014, JoshiNarayan2014, CordaNoInfoLoss, infoparadox2013} in nature. Henceforth, the convergence of such outcomes has launched an effective unification that begins to merge, generalize, and simplify an array of strictly thermal and non-strictly thermal quantization approaches to a single, consolidated approach of effective states that acknowledges further insight into the physical structure, behavior, and effects of BHs. 

First, we discussed numerous approaches that initiated universal upper and lower bounds on the area quanta for non-extremal BHs that emit or absorb particles, which may or may not be charged. We reviewed the mechanisms and predicted quanta for both uncharged and charged particles, along with the relevant aspects of wave-particle duality for the BH mass-energy spectrum. Therefore, we conveyed the importance of linking the discrete particles to continuous waves with perturbation field QNMs that encode the BH's asymptotic behavior of spectral states and transitions. Subsequently, we identified a series of damped harmonic oscillator QNM configurations and strictly thermal corrections that were systematically deployed to encode a BH's behavior and quantization of area and entropy. Next, we shifted to the strictly thermal spectrum deviation corrections \cite{Parikh-Wilczek2000} that inspired numerous crucial follow-up explorations \cite{CordaNoInfoLoss, rbanerjee, rbanerjee2, Majhi, Majhi2, rbanerjee3, Banerjee4} with a subsequent application of QNMs and effective states \cite{Corda-IJMPD-2012, Corda-AnnPhys-2013, CordaHendiKatebiSchmidt-JHEP-2013, CordaAHEP2013}.

In our opinion, the BH area and entropy quantization work that we chronologically reviewed in this paper highlights a series of striking scientific results that are beneficial for tackling the gigantic problems imposed by BHs in the domain of cutting-edge space-time physics. In the future, such findings should be subjected to additional rigorous analysis, debate, experimentation, and hard work via the scientific method. In particular, we suggest that future work should focus on applying the non-strictly thermal spectrum \cite{Parikh-Wilczek2000, rbanerjee, rbanerjee2, Majhi, Majhi2, rbanerjee3, Banerjee4} and QNM effective state framework \cite{CordaNoInfoLoss, Corda-IJMPD-2012, Corda-AnnPhys-2013, CordaHendiKatebiSchmidt-JHEP-2013, CordaAHEP2013} to additional classes of BHs and alternative unification approaches.

\section*{Acknowledgment}
We wish to thank the anonymous referees for the constructive criticisms and comments that enhanced the quality and application of this paper. N. O. Schmidt also wishes to thank his beloved wife Marissa, and his friends J. Dolifka and M. F. Boyle for their support and financial assistance with this research. S. H. Hendi thanks Shiraz University Research Council. The work of S. H. Hendi has been supported financially by Research Institute for Astronomy \& Astrophysics of Maragha (RIAAM), Iran.

\section*{Conflict of Interests}
The authors declare that there is no conflict of interests regarding the publication of this article.


\begin{thebibliography}{53}

\bibitem[1]{Hawking2014}S. W. Hawking, arXiv:1401.5761 [hep-th] (2014).

\bibitem[2]{Firewall2014}A. Almheiri, D. Marolf, J. Polchinski, and J. Sully, JHEP, 2, 62 (2013).

\bibitem[3]{JoshiNarayan2014}P. S. Joshi and R. Narayan, arXiv:1402.3055 [hep-th] (2014).

\bibitem[4]{CordaNoInfoLoss}C. Corda, arXiv:1304.1899 [gr-qc] (2013).

\bibitem[5]{infoparadox2013}B. Zhang, Q. Cai, M. Zhan, and L. You, Phys. Rev. D 87, 044006 (2013, First Award in the 2013 Awards for Essays on Gravitation).

\bibitem[6]{stringtheorybook}M. B. Green, J. H. Schwarz, and E. Witten. Superstring theory. Vol. 2. Cambridge university press (2012).

\bibitem[7]{LQGquantization}R. Gambini and J. Pullin, Phys. Rev. Lett. 110, 211301 (2013).

\bibitem[8]{LQGrelate}E. Frodden, M. Geiller, K. Noui, and A. Perez, arXiv:1212.4060 [gr-qc] (2013).

\bibitem[9]{LQGrelate2}A. Ghosh, K. Noui, and A. Perez, arXiv:1309.4563 [gr-qc] (2013).

\bibitem[10]{LQGrelate3}A. Corichi, J. Diaz-Polo, E. Fernandez-Borja, Phys. Rev. Lett. 98, 181301 (2007).

\bibitem[11]{LQGquant2009}C. Li, J. JiJian, and S. JiuQing, Science in China Series G: Physics, Mechanics and Astronomy 52.8, 1179-1182 (2009).

\bibitem[12]{su2cht2010}J. Engle, K. Noui, and A. Perez, Phys. Rev. Lett. 105, 031302 (2010).

\bibitem[13]{yukawa2013}M. Adeel Ajaib, I. Gogoladze, Q. Shafi, and C. Salih Un, JHEP 07, 139 (2013).

\bibitem[14]{E8}A. Garrett Lisi, arXiv:0711.0770 [hep-th] (2007).

\bibitem[15]{Hawking1975}S. W. Hawking, Commun. Math. Phys. 43, 199 (1975).

\bibitem[16]{Parikh-Wilczek2000}M. K. Parikh and F. Wilczek, Phys. Rev. Lett. 85, 5042 (2000).

\bibitem[17]{holoSuss}L. Susskind, Jour. Math. Phys. 36.11, 6377 (1995).

\bibitem[18]{holoEntropy}R. Emparan, JHEP 06, 012 (2006).

\bibitem[19]{originEntropy}J. D. Bekenstein, Phys. Rev. D 9, 3292 (1974).

\bibitem[20]{originEntropy2}A. Strominger and V. Cumrun, Phys. Lett. B 379, 99 (1996).

\bibitem[21]{originEntropy3}G. Horowitz, arXiv:gr-qc/9604051 (1996).

\bibitem[22]{spinfoamEntropy}J. M. Garcia-Islas, arXiv:0809.0304 (2008).

\bibitem[23]{spinfoamEntropy2}J. M. Garcia-Islas, arXiv:1303.2773 (2013).

\bibitem[24]{Bekenstein1972}J. D. Bekenstein, Lett. Nuovo Cim. 4, 737 (1972).

\bibitem[25]{Bekenstein1973}J. D. Bekenstein, Phys. Rev. D 7, 2333 (1973).

\bibitem[26]{Hod1999}S. Hod, Phys. Rev. Lett. 81 4293 (1998).

\bibitem[27]{Hod2008}S. Hod, Gen. Rel. Grav. 31, 1639 (1999, Fifth Award at Gravity Research Foundation).

\bibitem[28]{vagenasEPL2010}R. Banerjee, B. R. Majhi, E. C. Vagenas, Eur. Phys. Lett. 92, 20001 (2010).

\bibitem[29]{vagenasPRL2010}R. Banerjee, B. R. Majhi, E. C. Vagenas, Phys. Lett. B 686, 279 (2010).

\bibitem[30]{Maggiore2008}M. Maggiore, Phys. Rev. Lett. 100, 141301 (2008).

\bibitem[31]{vagenasPRL2011}B. R. Majhi, E. C. Vagenas, Phys. Lett. B 701, 623 (2011).

\bibitem[32]{rbanerjee}R. Banerjee, Phys. Lett. B 662, 62 (2008).

\bibitem[33]{rbanerjee2}R. Banerjee and B. R. Majhi, JHEP 06, 095 (2008).

\bibitem[34]{Majhi}B. R. Majhi, Phys. Rev. D 79, 044005 (2009).

\bibitem[35]{Majhi2}B. R. Majhi and S. Samanta, Ann. Phys. 325, 2410 (2010).

\bibitem[36]{rbanerjee3}R. Banerjee and B. R. Majhi, Phys. Lett. B 674, 218 (2009).

\bibitem[37]{Banerjee4}R. Banerjee, C. Kiefer, B. R. Majhi, Phys. Rev. D 82, 044013 (2010).

\bibitem[38]{Majhi3}B. R. Majhi, Phys. Lett. B 686, 49 (2010).

\bibitem[39]{gperez}G. P\'{e}rez-Nadal, Phys. Rev. D 77, 124033 (2008).

\bibitem[40]{Corda-IJMPD-2012}C. Corda, Int. Journ. Mod. Phys. D 21, 1242023 (2012, Honorable Mention at Gravity Research Foundation); C. Corda, JHEP 08, 101 (2011).

\bibitem[41]{Corda-AnnPhys-2013}C. Corda, Ann. Phys. 337, 49 (2013), latest version corrected by typos is available in arXiv:1305.4529v2.

\bibitem[42]{CordaHendiKatebiSchmidt-JHEP-2013}C. Corda, C. and S. H. Hendi and R. Katebi and N. O. Schmidt, JHEP 06, 08 (2013).

\bibitem[43]{CordaAHEP2013}C. Corda, C. and S. H. Hendi and R. Katebi and N. O. Schmidt, AHEP (Black Hole Special Issue), Article ID 527874 (2014).

\bibitem[44]{Banerjee-Majhi2009}R. Banerjee and B. R. Majhi, Phys. Lett. B 675, 243 (2009).

\bibitem[45]{Hawking1979}S. W. Hawking, ``The Path Integral Approach to Quantum Gravity'', in General Relativity: An Einstein Centenary Survey, eds. S. W. Hawking and W. Israel, Cambridge University Press (1979).

\bibitem[46]{Medvel-Vagenas2005}A. J. M. Medved and E. C. Vagenas, Mod. Phys. Lett. A20, 1723 (2005).

\bibitem[47]{Arzano-Medved-Vagenas2005}M. Arzano, A. J. M. Medved and E. C. Vagenas, JHEP 0509, 037 (2005).

\bibitem[48]{Vagenas2008}E. C. Vagenas, JHEP 0811, 073 (2008).

\bibitem[49]{Medved2008}A. J. M. Medved, Class. Quant. Grav. 25, 205014 (2008).

\bibitem[50]{CordaEPJC}C. Corda,  Eur. Phys. J. C 73, 2665 (2013).

\bibitem[51]{Banerjee-MajhiPLB}R. Banerjee and B. R. Majhi, Phys. Lett. B 674, 218 (2009).

\bibitem[52]{Shankaranarayanan}S. Shankaranarayanan, Mod. Phys. Lett. A 23, 1975 (2008).

\bibitem[53]{jzhang}J. Zhang, Phys. Lett. B 668, 353 (2008).

\bibitem[53]{gosh}A. Ghosh, P. Mitra, Phys. Rev. D 71, 027502 (2005). 

\end{thebibliography}
\end{document}